\documentstyle[12pt]{article}
\begin{document}

\begin{titlepage}
\vspace*{-4cm}
\begin{flushright}
EFI-97-22 \\
hep-ph/9705418
\end{flushright}
\vspace{1cm}
\begin{center}
\Huge Spontaneous CP Violation \\
\Huge in Models with Anomalous U(1)
\end{center}
\vspace{1cm}
\begin{center}
\large Merle Michael Robinson and Jacek Ziabicki
\end{center}
\begin{center}
\normalsize\it Enrico Fermi Institute and
the Department of Physics,\\
\normalsize\it University of Chicago, 5640 S. Ellis Ave.,
Chicago, IL 60637
\end{center}
\vspace{1cm}
\centerline{May 25, 1997}
\vspace{1cm}

\begin{abstract}

We examine a class of Froggatt-Nielsen models with an anomalous
$U(1)$ as the flavor dependent symmetry. Anomaly cancellation and
unbroken supersymmetry impose constraints on the $U(1)_X$ charges
of the fermions and the vacuum expectation values of the
symmetry-breaking scalars. We show by example that it is possible
to find models that reproduce the observed masses and mixings of
the standard model fermions, and exhibit a realistic amount of CP
violation.

\end{abstract}

\end{titlepage}

\section{Introduction}

Froggatt-Nielsen models \cite{fn} offer an elegant explanation for
the observed hierarchy in fermion masses and mixings. They have
recently attracted much interest in the context of string theory.
One hopes that a more fundamental theory can provide more details
on how the rather general Froggatt-Nielsen mechanism is
implemented, and give some predictions. One such attempt is to
assume that the requisite broken symmetry is the anomalous $U(1)$
of a compactified string theory \cite{ir}. The anomalies are
cancelled by the Green-Schwarz mechanism \cite{gs}, which imposes
constraints on the $U(1)_X$ charges of the standard model
particles.

In an earlier paper \cite{rz}, we examined such a model with the
additional assumption that supersymmetry must not be broken at the
string scale. In the simplest case where the $U(1)_X$ symmetry is
broken by the vacuum expectation value (VEV) of only one field,
this determines the symmetry breaking scale and hence the hierarchy
parameter $\lambda$. We found it was possible to find models
($U(1)_X$ charge assignments for the standard model particles) that
would produce phenomenologically viable masses and mixings, while
at the same time satisfying all the anomaly and supersymmetry
constraints.

Although the results were encouraging, our simplest model was not
rich enough to include CP violation. This could be achieved by
allowing the order one coefficients in the Yukawa mass matrices to
be complex; in this paper, we examine the case of spontaneous CP
violation --- we assume that the coefficients are real, but there
are two scalar fields breaking the the $U(1)_X$ symmetry. We find,
without any fine tuning or additional assumptions, a number of
models that satisfy all anomaly and supersymmetry constraints, and
include acceptable masses, mixings, and the right amount of CP
violation. The result does not strongly depend on the particular
choice of the VEVs of the symmetry-breaking fields or the
coefficients in the mass matrices. We conclude that the existence
of such models is a generic phenomenon, rather than an exception.
This allows more room for string model building, since the
underlying fundamental string theory may not produce complex
couplings.

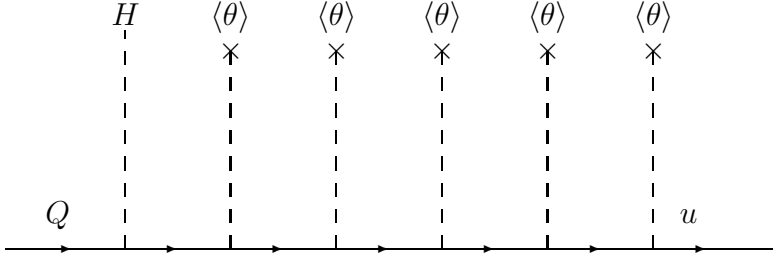
\begin{figure}
\begin{center}
\newsavebox\dashed
\begin{picture}(300,100)
\put(5,0){\line(1,0){290}}
\sbox{\dashed}{\multiput(0,0)(0,10){8}{\line(0,1){5}}%
\put(-4.58,72){$\times$}\put(-7.5,85){$\langle\theta\rangle$}}
\multiput(50,0)(0,10){8}{\line(0,1){5}}\put(50,80){\line(0,1){3}}
\put(45,85){$H$}
\multiput(90,0)(40,0){5}{\usebox\dashed}
\multiput(30,0)(40,0){7}{\vector(1,0){0}}
\put(20,10){$Q$}
\put(260,10){$u$}
\end{picture}
\end{center}
\caption{Effective Yukawa couplings in Froggatt-Nielsen models}
\label{fig1}
\end{figure}

Froggatt-Nielsen models are built on the assumption that a
flavor-dependent $U(1)$ gauge symmetry is broken by the VEV of a
field that is a singlet under the standard model gauge
interactions. Charge conservation disallows direct Yukawa couplings
of quarks and leptons; the interactions required by the standard
model between the up and down type quarks and leptons and the Higgs
boson proceed through higher-order tree diagrams such as the one
shown in Fig.~\ref{fig1}.
With all the three-point couplings of the same order of magnitude,
one can obtain the hierarchy of the standard model couplings
required by experiment if the VEV of the symmetry-breaking field
$\theta$ is slightly below the mass of the intermediate fermions
--- so that $\lambda = f_\theta \langle \theta \rangle / M \simeq
0.2$. We assume that the Yukawa coupling $f_\theta$ of the $\theta$
field is of order one; we will set it to one for simplicity. If
$q_\theta = -1$, the resulting effective Yukawa mass matrix is
\begin{equation}
\label{eqmatrix}
{\bf Y}_u=
f_u \,\lambda^{q_{H1}}
\left(
\begin{array}{ccc}
\lambda^{q_{Q1}+q_{u1}}
&\lambda^{q_{Q1}+q_{u2}}
&\lambda^{q_{Q1}+q_{u3}}
\\
\lambda^{q_{Q2}+q_{u1}}
&\lambda^{q_{Q2}+q_{u2}}
&\lambda^{q_{Q2}+q_{u3}}
\\
\lambda^{q_{Q3}+q_{u1}}
&\lambda^{q_{Q3}+q_{u2}}
&\lambda^{q_{Q3}+q_{u3}}
\end{array}
\right).
\end{equation}
Throughout this paper, $q_{Qi}$, $q_{ui}$, $q_{di}$, $q_{Li}$, and
$q_{ei}$ will denote the $X$ charges of the left-handed fields:
quark doublets, up-type antiquarks, down-type antiquarks, lepton
doublets and positrons ($i=1,2,3$\/ is the family index). We use
$q_Q = \sum_{i=1}^3 q_{Qi}$, etc.\ as an abbreviation for the sum
of charges over families, but $q_H = q_{H1}+q_{H2}$ is the sum of
the $X$ charges of the two Higgs doublets. With the freedom to
choose the flavor symmetry charges for all the standard model
fields, we can approximate the experimental results. Thus, very
small ratios of masses and mixings are explained as powers of a
not-so-small number $\lambda \simeq 0.2$.

The above explanation is attractive but by itself not satisfactory:
no theoretical principle determines $\langle \theta \rangle$ or
$M$, we only know that for the mechanism to work, their ratio must
be about 0.2. We have to assume the existence of some intermediate
heavy fermions that will carry color and hypercharge. Unless $M$ is
above the unification scale, those fermions will interfere with
gauge coupling unification.

The gauge groups of compactified string models often include an
anomalous $U(1)$ factor. Taking such a model as our underlying
theory brings a number of benefits. The energy scale of the problem
is around the string scale $M_s$, and the symmetry-breaking scale
should be somewhat lower, so having $M = M_s$ and $\langle \theta
\rangle = 0.2 M_s$ is quite acceptable. This avoids the problem
with gauge coupling unification. Cancelling the anomalies via the
Green-Schwarz mechanism imposes constraints on the otherwise
unrestricted Froggatt-Nielsen model. Requiring that supersymmetry
remain unbroken at the high scale gives another constraint, which
may be used to predict the value of the ratio $\langle \theta
\rangle / M$. Finally, it gives the Froggatt-Nielsen construction a
fundamental background by setting it in the context of string
theory.

\section{Supersymmetry and anomaly constraints}

In theories with anomalous $U(1)$, the $D$ term corresponding to
$U(1)_X$ is modified by a Fayet-Iliopoulos term \cite{ad,di,ds}:
\begin{equation}
\label{eqDterm}
D={g_s M_s^2\over192\pi^2} {\,\rm tr\,} Q + \sum_i q_i |\phi_i|^2,
\end{equation}
where $M_s$ is the string scale, $g_s$ is the string coupling
constant, and $\phi_i$ are all the scalars of the theory. In order
for supersymmetry to be left unbroken at high energies, we must
have $\langle D \rangle = 0$. Assuming that $\theta$ is the only
field that develops an expectation value, this determines the VEV
of $\theta$, and the hierarchy parameter $\lambda$, in terms of the
$X$ charges of all the fermions:
\begin{equation}
\label{eqsqrt}
\lambda =
{\langle\theta\rangle\over M_s} =
\sqrt{{-g_s\over192\pi^2} {{\,\rm tr\,} Q \over q_\theta}}\;.
\end{equation}
We are looking for models with $X$ charge assignments such as to
give $\lambda \simeq 0.2$.

We need the Green-Schwarz mechanism to cancel mixed anomalies of
$U(1)_X$ with the standard model gauge groups. The anomaly
coefficients are
\begin{eqnarray}
\label{eqC123}
C_1 &=& {1\over6} (q_Q + 8q_u + 2q_d + 3q_L + 6q_e + 3q_H)
\nonumber\\
C_2 &=& {1\over2} (3q_Q + q_L + q_H) \nonumber\\
C_3 &=& {1\over2} (2q_Q + q_u + q_d) \\
C_{\rm grav} &=& {1\over24}
\sum_{\mbox{\scriptsize all fields}}\!\!\!\!\! q_i \nonumber\\
&=& {1\over24}
(6q_Q + 3q_u + 3q_d + 2q_L + q_e + 2q_H + q_\theta + q_X) \nonumber
\end{eqnarray}
where $C_1 = {\,\rm tr\,}\left[Q (Y/2)^2\right]$ is the coefficient
of the $U(1)_X \, \left[U(1)_Y\right]^2$ anomaly, and $C_{2,3} =
{1\over2} {\rm tr}_{2,3}\,Q$ are the $U(1)_X
\left[SU(2)_L\right]^2$ and $U(1)_X \left[SU(3)_c\right]^2$
anomalies. (The trace is over fermions with $SU(2)_L$ and $SU(3)_c$
charges, respectively.) $C_{\rm grav}$ is the mixed gravitational
anomaly, and $C_X = {\,\rm tr\,} Q^3$ is the cubic anomaly
$\left[U(1)_X\right]^3$. In order for anomalies to be cancelled, we
must have \cite{i}
\begin{equation}
\label{eqcratio}
C_1:C_2:C_3:C_X:C_{\rm grav} = {5\over3}:1:1:1:1.
\end{equation}

In this paper, we are only interested in $X$ charges of the fields
that either are part of the standard model or break the $U(1)_X$
symmetry. We will refer to such charge assignments as ``models'' or
``examples''. Our framework allows any number of fields that are
singlets under the standard model gauge groups and do not break
$U(1)_X$. Except for (\ref{eqcratio}), we make no assumptions about
their $X$ charges. Although we cannot evaluate $C_X$ and $C_{\rm
grav}$ independently of Eq.~(\ref{eqcratio}), we can now write
$\lambda$ as a function of $C_3$, that is, in terms of the $X$
charges of the quark fields:
\begin{equation}
\label{eqsqrtc3}
\lambda =
\sqrt{{-g_s\over8\pi^2} {C_3 \over q_\theta}}\;.
\end{equation}

The order of magnitude of $C_3 / q_\theta$ can be estimated from
the determinants of quark mass matrices \cite{n}:
\begin{eqnarray}
\label{eqdet1}
\prod_{\mbox{\scriptsize all quarks}} \!\!\!\!\! m_q &=&
\left| \det{\bf Y}_u \right| \: \left| \det{\bf Y}_d \right| \\
& \sim & f_u^3 f_d^3 \,\lambda^{-(2q_Q + q_u + q_d +
3q_H)/q_\theta}.
\nonumber
\end{eqnarray}
If we take the mass ratios at the unification scale to be \cite{br}
\begin{eqnarray}
{m_u \over m_t} = {\cal O}(\lambda^8)\;; \qquad
{m_d \over m_b}&=&{\cal O}(\lambda^4)\;; \qquad
{m_e \over m_\tau}={\cal O}(\lambda^4) \nonumber\\
{m_c \over m_t} = {\cal O}(\lambda^4)\;; \qquad
{m_s \over m_b}&=&{\cal O}(\lambda^2)\;; \qquad
{m_\mu \over m_\tau} = {\cal O}(\lambda^2)\;,
\label{eqratios}
\end{eqnarray}
the product of the quark masses becomes
\begin{equation}
\label{eqlam}
\prod_{\mbox{\scriptsize all quarks}}
\!\!\!\!\! m_q \sim f_u^3 f_d^3\,\lambda^{18}.
\end{equation}
For $q_H=0$, Eqs.\ (\ref{eqdet1}) and (\ref{eqlam}) give \cite{br}
\begin{equation}
C_3 \simeq 9, \qquad \lambda \simeq 0.28\;.
\end{equation}

Another constraint we need to take into account comes from
$C_{YXX}$, the $\left[ U(1)_X \right]^2 \, U(1)_Y$ mixed anomaly.
$C_{YXX}$ depends only on the charges of the standard model fields,
and it cannot be cancelled by the Green-Schwarz mechanism, so for
every example we have to make sure that it vanishes:
\begin{eqnarray}
C_{YXX} &=& \!\!\!\!
\sum_{\mbox{\scriptsize all fields}} \!\!\!\!\! Y_i q_i^2
= \sum_{i=1}^3 \left(
q_{Qi}^2 - 2q_{ui}^2 + q_{di}^2 - q_{Li}^2 + q_{ei}^2
\right) \nonumber\\
&& {} - q_{H1}^2 + q_{H2}^2 = 0.
\end{eqnarray}

\section{Two symmetry-breaking fields}

In the case of one symmetry-breaking field, the VEV could always be
rotated by a gauge transformation so that it would be real. This is
no longer true for two symmetry-breaking fields with different $X$
charges. (Two fields with the same $X$ charges and all other
couplings will always appear together in the formulae and can be
treated as a single field $\theta = \theta_1 + \theta_2$.) We can
gauge away the imaginary part of one field, $\theta_1$, and are
left with three parameters: $|\langle \theta_1 \rangle|$, $|\langle
\theta_2 \rangle|$, and the angle between them, $\alpha$. The
powers of the complex VEV will make the Yukawa matrices and the CKM
matrix complex, and consequently may lead to  CP violation.

The anomaly conditions stay the same as in the model with one
$\theta$, but the supersymmetry condition becomes
\begin{equation}
q_{\theta1} {\left|\langle \theta_1 \rangle \right|^2 \over M_s^2}
+
q_{\theta2} {\left|\langle \theta_2 \rangle \right|^2 \over M_s^2}
+
{g_s\over 8 \pi^2} C_3 = 0
\end{equation}
and the element of the Yukawa mass matrix, for the excess charge $x
= q_{Qi}+q_{uj}$ is
\begin{equation}
\label{eqYuk12}
\left({\bf Y}_u\right)_{ij}=
\!\!\!\!\! \sum_{n_1 q_{\theta1} + n_2 q_{\theta2} = -x} \!\!\!\!\!
C(n_1,n_2)
{\langle \theta_1 \rangle^{n_1} \langle \theta_2 \rangle^{n_2}
\over M_s^{n_1+n_2}}
\end{equation}
where $C(n_1,n_2)$ is the combinatorial factor that in field theory
sums all the diagrams with $n_1$ insertions of $\theta_1$ and $n_2$
insertions of $\theta_2$. In the context of a specific string
model, it will be determined by a single string tree diagram, but
here we use the field-theory value
\begin{equation}
\label{eqCn1n2}
C(n_1,n_2) = {(n_1+n_2)! \over n_1! n_2!} f_\theta^{n_1+n_2}.
\end{equation}

We now have two distinct cases: one when the signs of $q_{\theta1}$
and $q_{\theta2}$ are the same (negative by our convention) and the
other when they are different. When the signs are the same, we can
roughly estimate $|\langle \theta_1 \rangle|$ and $|\langle
\theta_2 \rangle|$. There are two limit cases: when the VEV of one
field dominates ($|\langle \theta_1 \rangle| \gg |\langle \theta_2
\rangle|$ and $|\langle \theta_1 \rangle|^{1/|q_{\theta1}|} \gg
|\langle \theta_2 \rangle|^{1/|q_{\theta2}|}$), the dominant VEV
will be about the same as in a single-$\theta$ model with $\langle
\theta \rangle = |\langle \theta_1 \rangle|$. The other case is
when $|\langle \theta_1 \rangle| = |\langle \theta_2 \rangle|$.
Then,
\begin{equation}
{|\langle \theta_1 \rangle| \over M_s} =
\sqrt{{-g_s \over 8\pi^2}\,{C_3 \over q_{\theta1} + q_{\theta2}}}.
\end{equation}
In this case, $C_3$ can be estimated by following the same logic as
in the single-$\theta$ case (but with a greater margin of error).
If $|q_{\theta1}| < |q_{\theta2}|$, then the Yukawa matrix element
will be of order $\lambda^{n_1+n_2}$, where $n_1+n_2$ will be no
smaller than $(q_{Qi} + q_{uj})/ |q_{\theta2}|$. The product of the
determinants of mass matrices becomes
\begin{equation}
\prod_{\mbox{\scriptsize all quarks}}
\!\!\!\!\! m_q \sim f_u^3 f_d^3\,\lambda^{-2C_3/q_{\theta2}},
\end{equation}
and the resulting hierarchy parameter is
\begin{equation}
\lambda = \lambda_{\mbox{\scriptsize single-$\theta$}}
\sqrt{q_{\theta2} \over q_{\theta1} + q_{\theta2}}.
\end{equation}
For $\lambda_{\mbox{\scriptsize single-$\theta$}} = 0.28$,
$q_{\theta1} = -1$ and $q_{\theta2} = -2$, we get $\lambda = 0.23$.

Since the superpotential must be holomorphic in $\theta_1$ and
$\theta_2$, we must have $n_1, n_2 \ge 0$. This will give rise to
texture zeroes when the equation $n_1 q_{\theta1} + n_2 q_{\theta2}
= -x$ has no solution such that $n_1, n_2 \ge 0$.

The case where $q_{\theta1}$ and $q_{\theta2}$ have opposite signs
is different in many aspects. First, we have no easy way to
estimate the magnitude of the VEVs. If $|q_{\theta1}|$ and
$|q_{\theta2}|$ are relatively prime, the equation $n_1 q_{\theta1}
+ n_2 q_{\theta2} = -x$ always has infinitely many solutions, so
there are no texture zeroes. The sum (\ref{eqYuk12}) becomes an
infinite series which may or may not converge. In the following
section, we generate some numerical examples for the same-sign
case.

\section{Numerical results}

We now examine the two-theta model numerically through an
exhaustive search of all $X$ charge assignments for the standard
model fields, where the charges are in the range from $-10$ to 10.
(We adopt a normalization such that the charges will be integers.)
For the purpose of the search, we take the VEVs of the $\theta$
fields to be equal ($|\langle\theta_1\rangle| =
|\langle\theta_2\rangle|$), the angle between them $\alpha =
\pi/2$, and the charges $q_{\theta1} = -1$, $q_{\theta2} = -2$. We
use the tree-level value of the string coupling \cite{k,g} $$
g_s^2 = g_{GUT}^2/k_{GUT},
$$
with the Kac-Moody level $k_{GUT}=1$. The unified gauge coupling
constant $\alpha_{GUT}=g_{GUT}^2/4\pi \simeq 1/25$ gives $g_s
\simeq 0.7$. The Yukawa coupling $f_\theta = 1$ as before.

For each set of charges that satisfies the anomaly constraints, we
obtain the VEVs from the supersymmetry condition, then compute the
Yukawa mass matrices using Eq.\ (\ref{eqYuk12}), and diagonalize
them by singular value decomposition to obtain the masses and the
CKM matrix. We reject the examples where (a) any of the masses is
zero, or (b) two masses within the same sector are equal, e.g.\
$m_u = m_c$, or (c) the mass ratios are too far away from the
experimental values (\ref{eqratios}), or (d) the CKM matrix is too
different from the measured mixing matrix. To implement condition
(c), we introduce a ``badness'' score, for which a difference by a
factor of $0.22$ from a mass ratio in (\ref{eqratios}) is worth one
point; we sum those points for all the ratios (\ref{eqratios}). Any
example with badness greater than three, or more than two badness
points in any one sector, is rejected; our best examples have
badness around one. It should be noted that our method is not
dependent on the particular choice of the mass ratios. If at some
point a slightly different choice turns out to be better (e.g.\
because it solves another problem \cite{n}), it may change which
examples will be picked, but it will not affect our conclusions.
(We don't even need to write (\ref{eqratios}) in terms of powers of
$\lambda$: any set of mass ratios will do.)

For condition (d) we reject all examples where $|V_{12}|$ or
$|V_{21}|$ is not in the range 0.17--0.25. The Cabibbo angle is the
most precisely measured element of the CKM matrix and it is also
almost invariant when renormalized to the unification scale
\cite{op}. We will see in the examples below that the remaining CKM
matrix elements are within an order of magnitude of the
experimental values. The 90\% confidence experimental limits on the
magnitude of the CKM matrix elements \cite{rev}, renormalized to
the GUT scale \cite{op}, are
$$
\left(
\begin{array}{ccc}
0.9745\mbox{ to } 0.9757 &
0.219 \mbox{ to } 0.224  &
0.001 \mbox{ to } 0.003  \\
0.218 \mbox{ to } 0.224  &
0.9736\mbox{ to } 0.9750 &
0.023 \mbox{ to } 0.030  \\
0.002 \mbox{ to } 0.009  &
0.022 \mbox{ to } 0.032  &
0.9995\mbox{ to } 0.9997
\end{array}
\right).
$$

For each example we calculate $J^{CP}$, the invariant measure of CP
violation \cite{h}. $J^{CP}$ is defined as
\begin{equation}
J^{CP} = \left| {\rm Im}\, \left({\bf V}_{i\alpha} {\bf V}_{j\beta}
{\bf V}_{i\beta}^\ast {\bf V}_{j\alpha}^\ast \right) \right|
\end{equation}
(no summation, $i\not=j$, $\alpha\not=\beta$), which in the
Kobayashi-Maskawa parametrization of the CKM matrix \cite{km}
becomes
\begin{equation}
J^{CP} = c_1 c_2 c_3 s_1^2 s_2 s_3 \sin \delta,
\end{equation}
and in the Chau-Keung parametrization \cite{ck} used by the Review
of Particle Properties \cite{rev} it is
\begin{equation}
J^{CP} = c_{12} c_{13}^2 c_{23} s_{12} s_{13} s_{23} \sin
\delta_{13}.
\end{equation}

An important property of $J^{CP}$ is that it can be written in
terms of the absolute values of the CKM matrix elements:
\begin{eqnarray}
\left(J^{CP}\right)^2 &=& |{\bf V}_{ub}|^2 \, |{\bf V}_{cb}|^2 \,
|{\bf V}_{ud}|^2 \, |{\bf V}_{cd}|^2
\nonumber \\
&& {} - {1\over2} \left(1 - |{\bf V}_{ud}|^2 - |{\bf V}_{cd}|^2
- |{\bf V}_{ub}|^2 \right.\\
&& \left. {} + |{\bf V}_{ud}|^2 \, |{\bf V}_{cb}|^2 + |{\bf
V}_{ub}|^2 \,
|{\bf V}_{cd}|^2 \right)^2.
\nonumber
\end{eqnarray}
That means that if we can generate the correct magnitudes of the
CKM matrix elements, we will automatically generate the correct
amount of CP violation. Conversely, in a single-theta model without
CP violation \cite{rz}, we cannot correctly generate the small
elements of the CKM matrix. To be consistent with recent
measurements, $J^{CP}$ should be about $10^{-5}$ or less.

In order to generate the examples, we had to introduce a parameter,
the ``texture factor'' (TF). We have assumed that all the
coefficients in the Yukawa mass matrices are of order one; however,
setting them all to one does not produce acceptable examples. Since
there is no reason to believe that they are all equal to one, we
arbitrarily decided to multiply the (2,3), (3,2) and (3,3) entries
of ${\bf Y}_u$, and to divide the same entries of ${\bf Y}_d$, by
TF. (We chose to do it this way to avoid introducing many
additional parameters.)

We now proceed to the examples. With the $X$ charges
$$
\begin{tabular}{c|rrrrr}
$i$  &  $q_{Qi}$  &  $q_{ui}$  &  $q_{di}$  &  $q_{Li}$  &
$q_{ei}$  \\
\hline
1 &    8    &    9    &    0    &    3    &   10    \\
2 &    5    &    2    &  $-1$   &  $-3$   &    5    \\
3 &    1    &  $-1$   &  $-1$   &  $-6$   &    3
\end{tabular}
$$
we have $C_3=18$ and $\lambda_1=\lambda_2=0.23$.
With the texture factor ${\rm TF}=2$,
we find the fermion mass ratios (corresponding to badness 0.98)
\begin{eqnarray*}
\rule[-2pt]{0pt}{14pt}
{m_u \over m_t}      =   7.0\times10^{-6}, \;\:
{m_d \over m_b}     &=&  2.2\times10^{-3}, \;\:
{m_e \over m_\tau}   =   1.9\times10^{-3},\\
\rule{0pt}{18pt}
{m_c \over m_t}      =   2.1\times10^{-3}, \;\:
{m_s \over m_b}     &=&  2.6\times10^{-2}, \;\:
{m_\mu \over m_\tau} =   6.1\times10^{-2},\\
\rule[2pt]{0pt}{16pt}
{m_t \over f_u}      =   1.5             , \qquad\quad\;\;\:
{m_b \over f_d}     &=&  0.97            , \qquad\quad\;
{m_\tau \over f_d}   =   1.0             ,
\end{eqnarray*}
the CKM matrix (we show the absolute values of the elements)
\begin{equation}
{\bf V} = \left(
\begin{array}{ccc}
  0.97             & 0.24              & 3.6\times10^{-3} \\
  0.24             & 0.97              & 2.2\times10^{-2} \\
  8.9\times10^{-3} & 2.1\times10^{-2}  & 1.0
\end{array}
\right).
\end{equation}
and the CP violation invariant
\begin{equation}
J^{CP} = 1.2 \times 10^{-6}.
\end{equation}

While not all values of the texture factor work equally well (${\rm
TF} = 2$ is a good choice, and ${\rm TF} = 1$ is a poor one), we
were able to find examples for a range of TF between 1 and 3. This
shows that ${\rm TF}=2$ is not a necessary condition for the
existence of good examples. Here is one for ${\rm TF} = 1.8$:
$$
\begin{tabular}{c|rrrrr}
$i$  &  $q_{Qi}$  &  $q_{ui}$  &  $q_{di}$  &  $q_{Li}$  &
$q_{ei}$  \\
\hline
1 &    9    &   10    &    0    &    0    &    8    \\
2 &    5    &    3    &    0    &  $-1$   &    7    \\
3 &    0    &    0    &  $-2$   &  $-2$   &    0
\end{tabular}
$$
gives $C_3=19.5$ and $\lambda_1=\lambda_2=0.24$, the fermion mass
ratios (badness 0.85) are
\begin{eqnarray*}
\rule[-2pt]{0pt}{14pt}
{m_u \over m_t}      =   6.0\times10^{-6}, \;\:
{m_d \over m_b}     &=&  1.6\times10^{-3}, \;\:
{m_e \over m_\tau}   =   3.5\times10^{-3},\\
\rule{0pt}{18pt}
{m_c \over m_t}      =   2.3\times10^{-3}, \;\:
{m_s \over m_b}     &=&  5.9\times10^{-2}, \;\:
{m_\mu \over m_\tau} =   5.5\times10^{-2},\\
\rule[2pt]{0pt}{16pt}
{m_t \over f_u}      =   1.8             , \qquad\quad\;\;\:
{m_b \over f_d}     &=&  1.1             , \qquad\quad\;\;\:
{m_\tau \over f_d}   =   1.0             ,
\end{eqnarray*}
the CKM matrix (absolute values) is
\begin{equation}
{\bf V} = \left(
\begin{array}{ccc}
  0.98             & 0.22              & 2.9\times10^{-3} \\
  0.22             & 0.98              & 8.4\times10^{-3} \\
  1.2\times10^{-3} & 8.8\times10^{-3}  & 1.0
\end{array}
\right).
\end{equation}
and the CP violation invariant
\begin{equation}
J^{CP} = 1.5 \times 10^{-6}.
\end{equation}

To see that the results do not depend on the way the texture
factors were introduced, we ``randomly'' picked by hand the
coefficients of the up and down mass matrices. For the following
choice of coefficients
$$
{\bf Y}_u \sim
\left(
\begin{array}{rrr}
1.2  &  1.1  & -0.7 \\
0.95 &  1.5  & -2.0 \\
1.6  &  0.5  &  1.2
\end{array}
\right),
\quad
{\bf Y}_d \sim
\left(
\begin{array}{rrr}
0.8 &  -0.9  & -1.3 \\
0.9 &   1.4  &  2.0 \\
2.0 &   0.7  &  0.9
\end{array}
\right)
$$
we get
$$
\begin{tabular}{c|rrrrr}
$i$  &  $q_{Qi}$  &  $q_{ui}$  &  $q_{di}$  &  $q_{Li}$  &
$q_{ei}$  \\
\hline
1 &    7    &    5    &    1    &   10    &   10    \\
2 &    4    &    4    &    0    &  $-3$   &    8    \\
3 &    0    &    0    &    0    &  $-8$   &  $-5$
\end{tabular}
$$
$C_3=16$ and $\lambda_1=\lambda_2=0.22$. The fermion mass ratios
(badness 2.33) are
\begin{eqnarray*}
\rule[-2pt]{0pt}{14pt}
{m_u \over m_t}      =   5.6\times10^{-6}, \;\:
{m_d \over m_b}     &=&  5.0\times10^{-3}, \;\:
{m_e \over m_\tau}   =   2.5\times10^{-3},\\
\rule{0pt}{18pt}
{m_c \over m_t}      =   1.0\times10^{-2}, \;\:
{m_s \over m_b}     &=&  1.1\times10^{-1}, \;\:
{m_\mu \over m_\tau} =   3.1\times10^{-2},\\
\rule[2pt]{0pt}{16pt}
{m_t \over f_u}      =   1.2             , \qquad\quad\;\;\:
{m_b \over f_d}     &=&  1.2             , \qquad\quad\;\;\:
{m_\tau \over f_d}   =   1.0             ,
\end{eqnarray*}
the CKM matrix
\begin{equation}
{\bf V} = \left(
\begin{array}{ccc}
  0.98             & 0.21              & 2.5\times10^{-3} \\
  0.21             & 0.98              & 6.9\times10^{-2} \\
  1.7\times10^{-2} & 6.7\times10^{-2}  & 1.0
\end{array}
\right).
\end{equation}
and
\begin{equation}
J^{CP} = 1.5 \times 10^{-5}.
\end{equation}

We also looked for plausible examples in the case where the VEVs of
the two $\theta$ fields are not equal. For the ratio
$|\langle\theta_2\rangle| / |\langle\theta_1\rangle| = 0.1$, the CP
violation parameter was very small. For $|\langle\theta_2\rangle| /
|\langle\theta_1\rangle| = 0.9$, we have
$$
\begin{tabular}{c|rrrrr}
$i$  &  $q_{Qi}$  &  $q_{ui}$  &  $q_{di}$  &  $q_{Li}$  &
$q_{ei}$  \\
\hline
1 &    9    &   10    &  $-1$   &    2    &   10    \\
2 &    5    &    1    &  $-1$   &  $-3$   &    8    \\
3 &    1    &  $-1$   &  $-2$   &  $-8$   &    2
\end{tabular}
$$
$C_3=18$, $\lambda_1= 0.247$, and $\lambda_2=0.222$. With the
texture factor ${\rm TF}=2$ again, the mass ratios (badness 1.12)
are
\begin{eqnarray*}
\rule[-2pt]{0pt}{14pt}
{m_u \over m_t}      =   5.6\times10^{-6}, \;\:
{m_d \over m_b}     &=&  2.7\times10^{-3}, \;\:
{m_e \over m_\tau}   =   3.2\times10^{-3},\\
\rule{0pt}{18pt}
{m_c \over m_t}      =   4.9\times10^{-3}, \;\:
{m_s \over m_b}     &=&  5.6\times10^{-2}, \;\:
{m_\mu \over m_\tau} =   5.9\times10^{-2},\\
\rule[2pt]{0pt}{16pt}
{m_t \over f_u}      =   2.1             , \qquad\quad\;\;\:
{m_b \over f_d}     &=&  1.1             , \qquad\quad\;\;\:
{m_\tau \over f_d}   =   1.0             ,
\end{eqnarray*}
the CKM matrix
\begin{equation}
{\bf V} = \left(
\begin{array}{ccc}
  0.97             & 0.23              & 5.1\times10^{-3} \\
  0.23             & 0.97              & 1.3\times10^{-2} \\
  2.1\times10^{-3} & 1.4\times10^{-2}  & 1.0
\end{array}
\right).
\end{equation}
and
\begin{equation}
J^{CP} = 3.3 \times 10^{-6}.
\end{equation}

\section{Conclusions}

We have attempted to show that string-inspired Froggatt-Nielsen
models can be easily extended to include CP violation. We examined
compactified string models with anomalous $U(1)$ with anomalies
cancelled by the Green-Schwarz mechanism.

Two scalar fields breaking the $U(1)_X$ symmetry are needed to
spontaneously break CP. The assumption that supersymmetry remains
unbroken down to low energies leads to an estimate for the VEVs of
the two fields and consequently for the Froggatt-Nielsen hierarchy
parameters $\lambda_{1,2}$. The requirement of anomaly cancellation
puts constraints on the $X$ charges of the standard model fields.

We are not working within a specific string model. We focus on
model-independent features with very few important assumptions:
that the $U(1)$ symmetry in the Froggatt-Nielsen mechanism is
anomalous, that it comes from string theory, that it is broken by
the VEVs of two scalar fields, and that the coefficients of the
powers of $\lambda$ in the Yukawa mass matrices are real and of
order unity. We also assume that $f_\theta$, the Yukawa coupling of
the $\theta$ fields, is of order unity.

In the numerical computations we have introduced many unimportant
assumptions, such as the values of the texture factors, the charges
of the $\theta$ fields, the angle between their VEVs and the ratio
of their magnitudes. We have also set $f_\theta=1$ and used the
field-theory expression (\ref{eqCn1n2}) for $C(n_1,n_2)$. The
numbers in the examples depend on those input values, but the
qualitative features such as CP violation, masses and mixings in
rough agreement with experiment do not. The numerical values of the
mass ratios, mixings and the CP violation parameter can be
meaningfully calculated only for a specific string model, where the
unimportant assumptions will become unnecessary.

\section*{Acknowledgements}

We are grateful to Jeff Harvey and Joe Lykken for many useful
discussions. We would also like to thank George Hockney for help
with
optimizing our code. This research was supported in part by
NSF Grant No.\ PHY-9123780.

\end{document}